\documentclass[conference,a4paper]{IEEEtran}
\IEEEoverridecommandlockouts
\usepackage{cite}
\usepackage{amsmath,amssymb,amsfonts}
\usepackage{algorithmic}
\usepackage{graphicx}
\usepackage{textcomp}
\usepackage{xcolor}
\usepackage[caption=false,font=footnotesize]{subfig}
\usepackage{multirow}

\IEEEoverridecommandlockouts

\newcommand{\IEEEcopyrighttext}{%
  \textcopyright~2026 IEEE. 
  Personal use of this material is permitted. Permission from IEEE must be obtained for all other uses, in any current or future media, including reprinting/republishing this material for advertising or promotional purposes, creating new collective works, for resale or redistribution to servers or lists, or reuse of any copyrighted component of this work in other works.
}

\IEEEpubid{%
  \makebox[\columnwidth][l]{%
    \parbox[t]{\columnwidth}{%
      \raggedright
      \fontsize{7}{8}\selectfont
      \IEEEcopyrighttext
    }%
  }%
  \hspace{\columnsep}%
  \makebox[\columnwidth]{}%
}

\def\BibTeX{{\rm B\kern-.05em{\sc i\kern-.025em b}\kern-.08em
    T\kern-.1667em\lower.7ex\hbox{E}\kern-.125emX}}
\begin{document}

\title{A Low-Power Sparse Convolution Accelerator with Idle-First-Task-Assignment for Edge Vision\\

}
\author{
\IEEEauthorblockN{
Jingyue Zhuge\IEEEauthorrefmark{1}\IEEEauthorrefmark{2},
Johannes Partzsch\IEEEauthorrefmark{1},
Christian Mayr\IEEEauthorrefmark{1}\IEEEauthorrefmark{3}
}
\IEEEauthorblockA{\IEEEauthorrefmark{1}Chair of Highly-Parallel VLSI-Systems and Neuro-Microelectronics, TUD Dresden University of Technology, Germany}
\IEEEauthorblockA{\IEEEauthorrefmark{2}Center for Scalable Data Analytics and Artificial Intelligence (ScaDS.AI Dresden)}
\IEEEauthorblockA{\IEEEauthorrefmark{3}Centre for Tactile Internet with Human-in-the-loop (CeTI)}
\IEEEauthorblockA{Email: jingyue.zhuge@tu-dresden.de}
}


\maketitle

\begin{abstract}
In recent years, edge-vision monitoring systems for applications such as smart animal husbandry have faced strict tripartite constraints: maintaining input resolution under extremely limited transmission bandwidth and strict power budgets. Conventional dense convolutional neural networks (CNNs) cannot satisfy the resource limits of such constrained IoT nodes. To address this challenge, this paper presents a low-power sparse convolution accelerator for edge devices, fabricated and validated in a 16 nm process. First, the accelerator adopts a bitmap-based format for compression in both data transmission and computation, effectively reducing memory and bandwidth overhead. Second, to mitigate load imbalance in sparse computation, an Idle-First-Task-Assignment (IFTA) dynamic scheduling strategy is proposed, significantly reducing processing-element (PE) idle time and improving multiplier utilization. In addition, a dedicated dataflow is designed to support and accelerate depthwise separable convolution (DWConv), which is widely used in lightweight networks. Experimental results show that the chip occupies only 0.5~mm$^2$ core area and consumes as little as 12--16~mW. On ImageNet, for sparse VGG16 and MobileNetV2, the proposed accelerator achieves 6.5$\times$ and 2.8$\times$ speedups, respectively, over traditional dense accelerators, and also delivers significant performance gains over the existing sparse accelerator.
\end{abstract}

\begin{IEEEkeywords}
edge AI accelerator, sparse convolution, low power, dynamic scheduling, depthwise separable convolution
\end{IEEEkeywords}

\section{Introduction}
\label{sec:introduction}

With the widespread deployment of IoT devices, using visual sensors to monitor the feeding and mobility status of livestock has become increasingly prevalent. These vast pasture environments, typically lacking reliable power grids and high-speed network coverage, force IoT monitoring nodes to rely on strictly constrained power sources and low-bandwidth communication protocols. At the same time, capturing small individual animals across broad backgrounds dictates a relatively high input image resolution to maintain recognition accuracy. Since the slow behavioral changes of livestock permit a low processing frame rate (typically around 1 FPS), the primary challenge for edge vision hardware lies in balancing guaranteed input resolutions with low transmission bandwidth and stringent power constraints.

Processing these high-resolution images rapidly exhausts on-chip SRAM capacity, particularly during the initial CNN layers where channel expansion generates massive intermediate feature maps. While patch-wise inference (e.g., MCUNetV2~\cite{lin2021memory}) ieffectively addresses this memory bottleneck by dividing the input into independently processed regions, it inevitably incurs additional computational overhead to handle overlapping areas.

To further alleviate communication bandwidth demands, split computing has emerged as a promising paradigm. In this approach, initial CNN layers are executed at the edge endpoint, transmitting only the extracted, down-sampled feature maps to the cloud. While processing more layers locally reduces transmission payloads, it significantly increases the computational complexity and power consumption at the edge. Although previous sparse convolution accelerators~\cite{han2016eie, parashar2017scnn,chen2017eyeriss,chen2019eyeriss, gondimalla2019sparten,yang2020procrustes,zhang2020snap,deng2021gospa,wang2022paca,ding2023optimized,choi2023efficient,meng2023efficient,yang2023isosceles,yao2024eyelet,wu2024edge,guo2024efficient,yang2024fpga,liu2024efficient,chen2025gupa} attempt to mitigate this by exploiting zero-valued data, they frequently struggle with processing element (PE) load imbalance, which severely limits performance improvements. Moreover, they largely lack tailored optimizations for depthwise separable convolutions (DWConv)—a staple operation in lightweight edge networks like MobileNetV2~\cite{sandler2018mobilenetv2}.

To address these critical bottlenecks, we propose a low-power sparse convolution accelerator. It leverages Bitmap format to compress data to minimize transmission volumes and exploits the sparsity of activations and weights to bypass ineffectual computations, thereby reducing power consumption. Concurrently, we propose  Idle-First-Task-Assignment strategy~(IFTA) to fundamentally resolve the PE load imbalance issue. To accommodate efficient CNN architectures, a specialized dataflow is also designed to fulfill the unique computational requirements of DWConv.

The main contributions of this paper are as follows:
\begin{itemize}
    \item \textbf{Utilizing sparse convolution combined with patch-wise inference} to balance the additional computational overhead and peak memory requirements under severely memory-constrained conditions.
    \item \textbf{Proposing a multi-dimensional Idle-First-Task-Assignment strategy} to resolve the load imbalance problem among PEs, significantly improving PE utilization.
    \item \textbf{Designing a dataflow} to efficiently map and accelerate the computational demands of DWConv.
\end{itemize}

\section{Hardware Architecture}
\label{sec:arch}
\begin{figure}[htbp]
    \centerline{\includegraphics[width=0.8\linewidth]{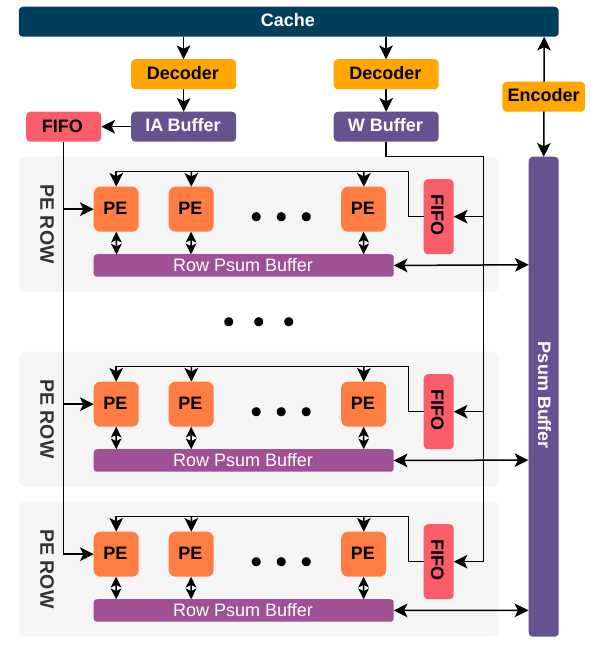}}
    \caption{Architecture overview.}
    \label{fig:arch}
\end{figure}

The overall architecture of our accelerator is shown in Fig. \ref{fig:arch}. The top-level architecture consists of an on-chip cache (Cache), a dual-way bitmap decoder (Decoder), multi-level buffers for input activations, weights, and partial sums, and a two-dimensional processing unit array composed of multiple PE rows (PE ROW). Each PE contains a MAC unit, as well as buffers for input activations and weights. Each PE row contains multiple PEs, and each PE row has a FIFO and a row partial sum buffer.

\subsubsection{Decoder}
\label{sec:decoder}
\begin{figure}[htbp]
    \centerline{\includegraphics[width=1\linewidth]{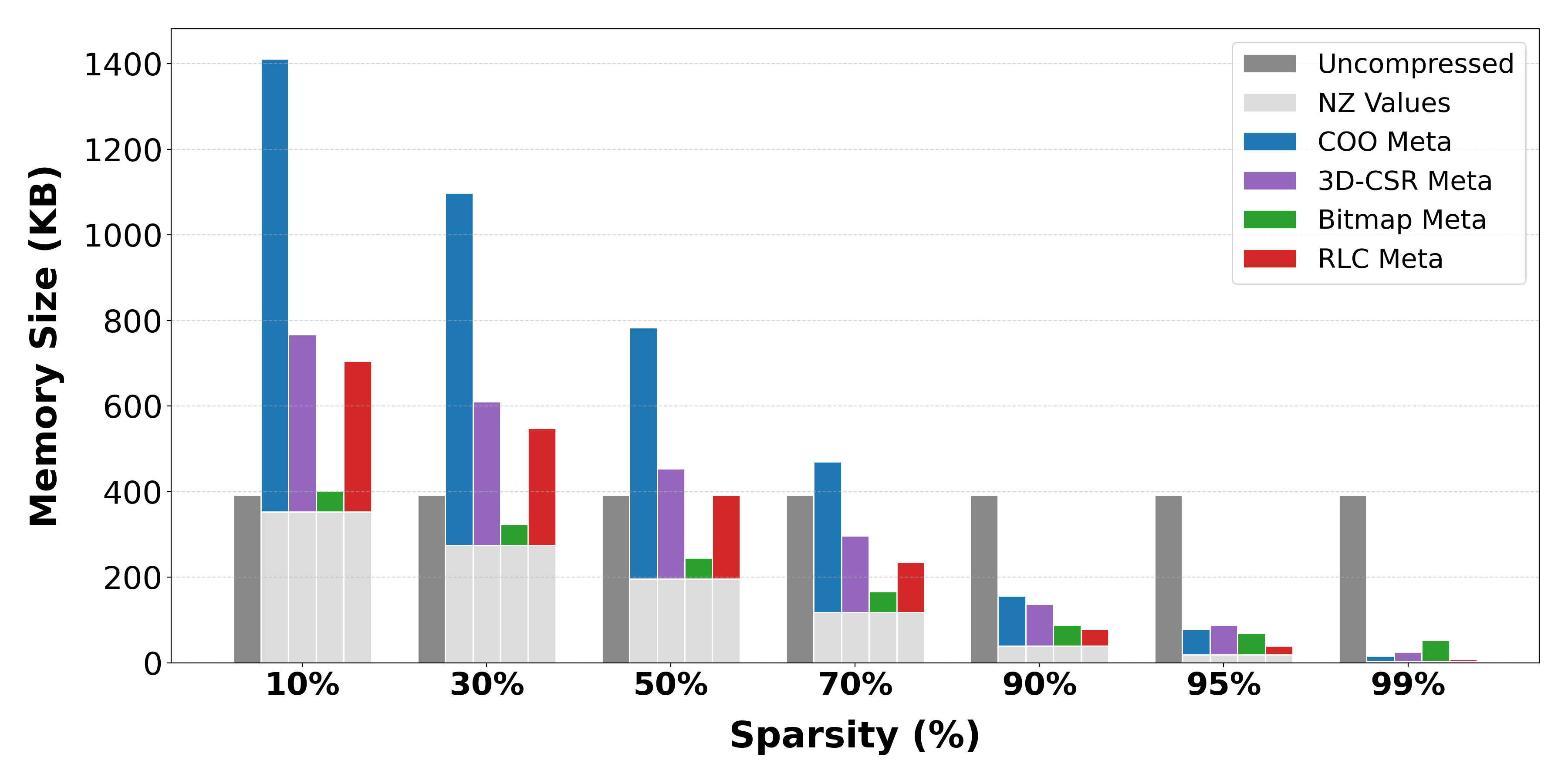}}
    \caption{Memory footprint breakdown of various sparse tensor formats (Uncompressed, COO, 3D-CSR, Bitmap, and RLC). The evaluation is performed on a $112 \times 112 \times 32$ tensor with \texttt{int8} precision across different sparsity levels (10\% to 90\%). The stacked bars distinguish the shared baseline of non-zero (NZ) values from the specific metadata overhead introduced by each scheme.}
    \label{fig:spformat}
\end{figure}

To achieve efficient data compression/decompression and effective non-zero value matching, a sparse compression format has to be applied. Fig. \ref{fig:spformat} illustrates the memory footprint of different sparse tensor formats (Uncompressed, COO, 3D-CSR, Bitmap, and RLC). It is evident that the Bitmap format generates the least metadata across low to high sparsity levels. When the sparsity exceeds 90\%, the metadata of Bitmap already occupies more memory than the non-zero values, This means that, compared to other formats, the bitmap format performs poorly at high sparsity levels; for example, the RLC format has a smaller memory footprint than the bitmap format. In the case of 99\% sparsity, CSR and COO are more efficient. However, the non-zero value matching efficiency of RLC format is relatively low, making it unsuitable for sparse accelerators. The boundary case of sparsity greater than 99\% is exceptionally rare in CNNs, so we choose the Bitmap format as our sparse data representation method. 

The decoder supports two modes: pass-through and decode mode. In pass-through mode, the decoder directly compacts the non-zero values read from the cache and passes them to the corresponding buffers. In this mode, data can be continuously read from the cache without waiting for decoding, thus maximizing the utilization of cache bandwidth. In decode mode, the decoder needs to decode according to the Bitmap, inserting zero values for non-zero values, and then transmitting them to the buffers vector by vector. In the case of sparsity, for example, when loading 32 non-zero values from SRAM, it may require multiple vector decoding transmissions to consume these non-zero values. The transmission speed in decode mode is slower, but it can fully unpack the vectors, making it more suitable for DW convolution, as detailed in \ref{sec:dataflow}.

\subsubsection{Idle-First-Task-Assignment (IFTA) for Load Balancing}
\label{sec:IFTA}
\begin{figure}[htbp]
    \centering
    \subfloat[Load Imbalance]{\includegraphics[width=0.7\linewidth]{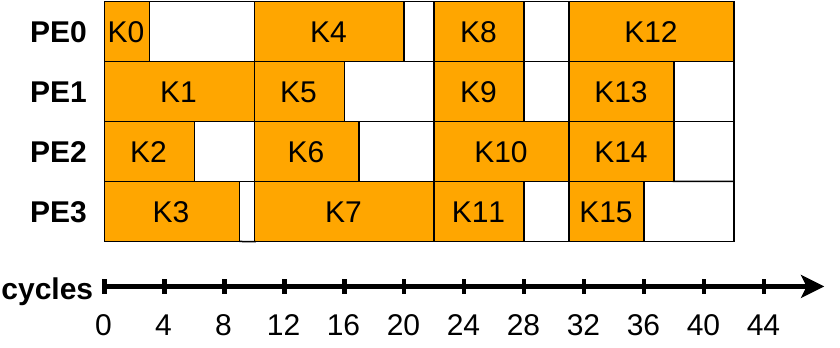}}
    \\
    \subfloat[IFTA]{\includegraphics[width=0.7\linewidth]{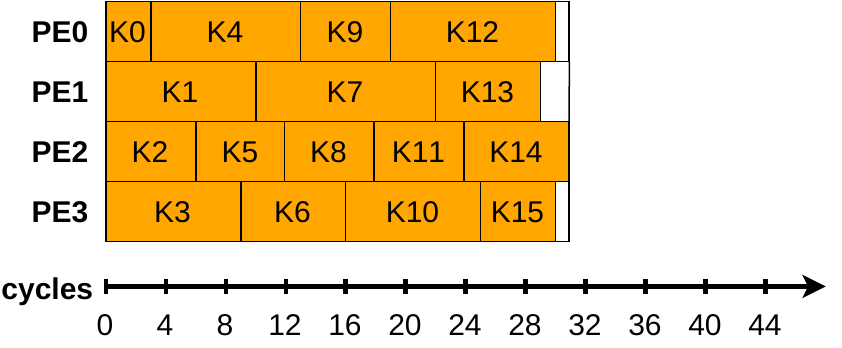}}
    \caption{Load balancing comparison between (a)load imbalance and (b)our proposed IFTA method.}
    \label{fig:IFTA_compare}
\end{figure}
Load imbalance is a common and challenging issue in sparse accelerators, leading to underutilization of multipliers and longer execution times. We propose a load balancing method called Idle-First-Task-Assignment (IFTA) to address this problem. As shown in the example in Fig. \ref{fig:IFTA_compare}, with a total of 16 convolution kernels, traditional accelerators need to wait for all PEs to complete their computations before executing the next round of operations. In contrast, the IFTA method immediately assigns the next round of data to a PE as soon as it completes its computation, allowing it to continue processing without waiting for other PEs. For instance, after PE0 completes K0, it directly starts processing K4, and after PE2 completes K2, it directly starts processing K5. It can be observed that the number of convolution kernels executed by each PE can vary, with one PE executing 5 convolution kernels while another executes only 3. The IFTA method can significantly reduce idle time and improve multiplier utilization, thereby shortening execution time by approximately 25\%.

\begin{figure}[htbp]
    \centerline{\includegraphics[width=0.8\linewidth]{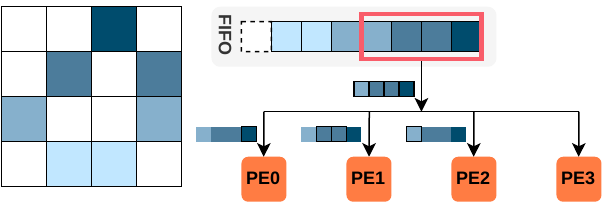}}
    \caption{Data distribution}
    \label{fig:data_dist}
\end{figure}
To implement our IFTA method, we introduce a FIFO in each PE row. The weight buffer broadcasts data simultaneously to the FIFOs of all PE rows. This avoids simultaneous access to different weight data in the weight buffer by each row. The FIFO also serves as a buffer to balance the task processing speed among the PE rows, and we configure its depth to be consistent with the number of PEs in each row.
We also handle the data distribution from the FIFO to the PEs in a special way. When the data is very dense, meaning that the output word width of the FIFO is insufficient to encompass all non-zero values of two weight vectors, the FIFO operates like a regular FIFO, outputting only a compact weight vector containing only non-zero values. Although this requires multiple cycles to complete the transmission for multiple PEs, when the data is dense, the computation time of the PEs can effectively cover the data transmission time of the FIFO, forming a pipeline without causing idle time for the PEs. When the data is very sparse, meaning that the output word width of the FIFO is sufficient to encompass all non-zero values of multiple weight vectors, the FIFO will transmit to multiple PEs at once, along with corresponding data offsets. As shown in Fig. \ref{fig:data_dist}, the FIFO simultaneously transmits data to PE0, PE1, and PE2, which only receive non-zero values corresponding to their respective weight vectors. This can reduce the number of cycles for data transmission in sparse cases. If we were to continue transmitting only one weight vector to one PE at a time in sparse conditions as we do in dense conditions, it would lead to data transmission becoming the biggest bottleneck due to the extremely short computation cycles of each PE. Our method effectively addresses this issue.

For input activation vectors, we use the same strategy to distribute them among PE rows to achieve multi-dimensional load balancing. To prevent PSUM write-back from becoming a bottleneck, we use a row partial sum buffer in each PE row to support simultaneous reading or writing back by all PEs in a row. After a PE row completes its weight iteration tasks, the row PSUM buffer will write back to the PSUM buffer and load the next round of PSUMs that need to be loaded. We implement the IFTA method through our data distribution approach and the design of the row partial sum buffer.


\subsubsection{Tile-based Processing} 
\label{sec:dataflow}

Our accelerator processes data based on tiles. We process the multiplication of an input activation matrix and a weight matrix at a time, specifically $ia[h,w:w+tile\_w,ci:ci+tile\_ci] \times w[ci:ci+tile\_ci,co:co+tile\_co,r,s]$. To maximize the advantages of the IFTA method, we set the number of weight vectors to be four times the number of PEs in each row, i.e., $tile\_co=4*N_{PE}$. All PEs within a PE row share the same IA vector and perform IFTA on the weight vectors, while between PE rows, we use a shared weight matrix approach and perform IFTA on the IA vectors.

\textbf{Depthwise separable convolution}: The computation of sparse depthwise separable convolution presents unique mapping challenges. In DW convolution, each output channel corresponds to only one input channel, so the standard convolution data flow would lead to each PE processing only one weight scalar, which completely loses the computational acceleration advantage brought by sparsity. To solve this problem, we adopt a different data flow approach in DW convolution. We swap the buffers of IA and W, allowing each PE row to share a weight vector of length $K$, where $K$ is the size of the convolution kernel, and different rows use weight vectors corresponding to different filters. The iteration within the PE row uses the IA vector, which is equivalent to performing a 1D convolution with a kernel of size $K$ on the vector $ia[h,w:w+tile\_w,ci]$. The distribution of weight vectors between PE rows still uses the aforementioned distribution method. Due to the very small vector length, it can be transmitted to multiple PE rows at once to reduce the number of cycles for data transmission. For IA vectors, we need to use the decode mode of the decoder to expand the compact non-zero values into complete vectors to meet the requirements of sliding windows. Moreover, we cannot use broadcasting to transmit to all PE rows' FIFOs because different PE rows need to use IA vectors corresponding to their respective filters. However, since the sliding window moves only by stride elements at a time, i.e., $offset1=offset0+stride$, all can be transmitted at once from FIFO to PE. Therefore, despite the very small vector length, efficient computation can still be performed in IFTA mode.

\section{Evaluation and results}
\label{sec:result}

To evaluate our accelerator, we built a cycle-accurate accelerator simulator for a dense accelerator (TPU-like systolic array)~\cite{jouppi2017datacenter}, SparTen~\cite{gondimalla2019sparten}, and our accelerator. All three accelerators use the same configuration.
Table \ref{tab:hardware-parameters} lists the hardware parameters of our accelerator. The mask length processed by each PE is set to 32. SparTen also exploits both activation and weight sparsity and uses a software-hardware co-design approach called Greedy Balancing to address load imbalance. The simulator can faithfully capture any idling caused by memory access and load imbalance. 
We also performed a complete RTL design and taped out using TSMC's 16nm technology.

\begin{table}[htbp]
\caption{Hardware Parameters}
\begin{center}
\begin{tabular}{|l|c|}
\hline
\textbf{Parameter} & \textbf{Value} \\
\hline
PEs per PE Row & 8 \\
\hline
Number of PE Rows & 8 \\
\hline
Row PSUM Buffer & 128 B \\
\hline
Cache Size & 2 MB, 4 Banks, 32B line \\
\hline
IA Buffer Size & 1 KB \\
\hline
W Buffer Size & 1 KB \\
\hline
PSUM Buffer Size & 4 KB \\
\hline
MAC Units per PE & 1 \\
\hline
Bitmap length per PE & 32 \\
\hline
\end{tabular}
\label{tab:hardware-parameters}
\end{center}
\end{table}

\subsubsection{Performance evaluation}
\label{sec:performance}
We use the pruning method from~\cite{han2015deep} to perform unstructured pruning on VGG16~\cite{simonyan2014very} and MobilenetV2~\cite{sandler2018mobilenetv2}, and fine-tune and evaluate them on the ImageNet~\cite{deng2009imagenet} dataset using Pytorch. The final sparsities, excluding fully connected layers, are 68\% and 35\%, respectively, while maintaining accuracy. 
\begin{figure}[htbp]
    \centering
    \subfloat[VGG16 Speedup]{\includegraphics[width=1\linewidth]{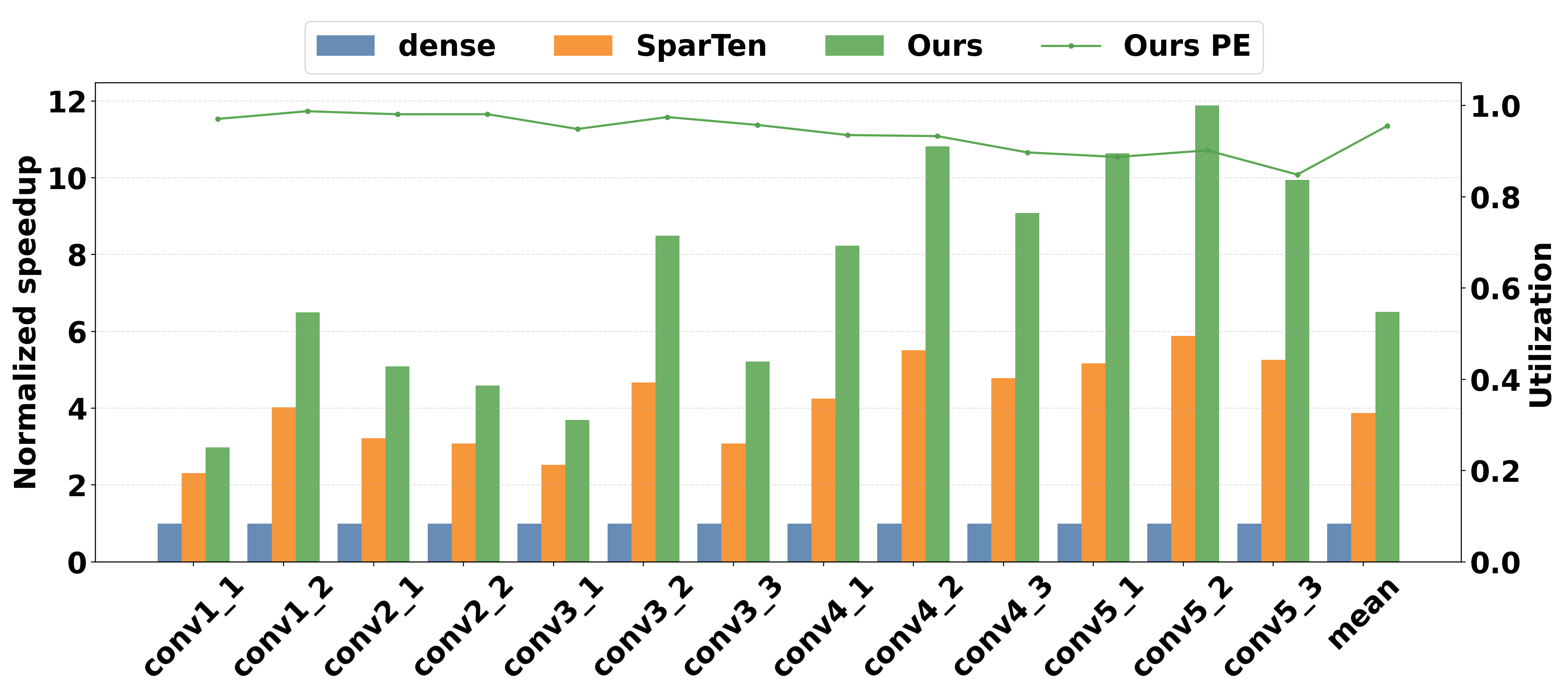}}
    \\
    \subfloat[MobilenetV2 Speedup]{\includegraphics[width=1\linewidth]{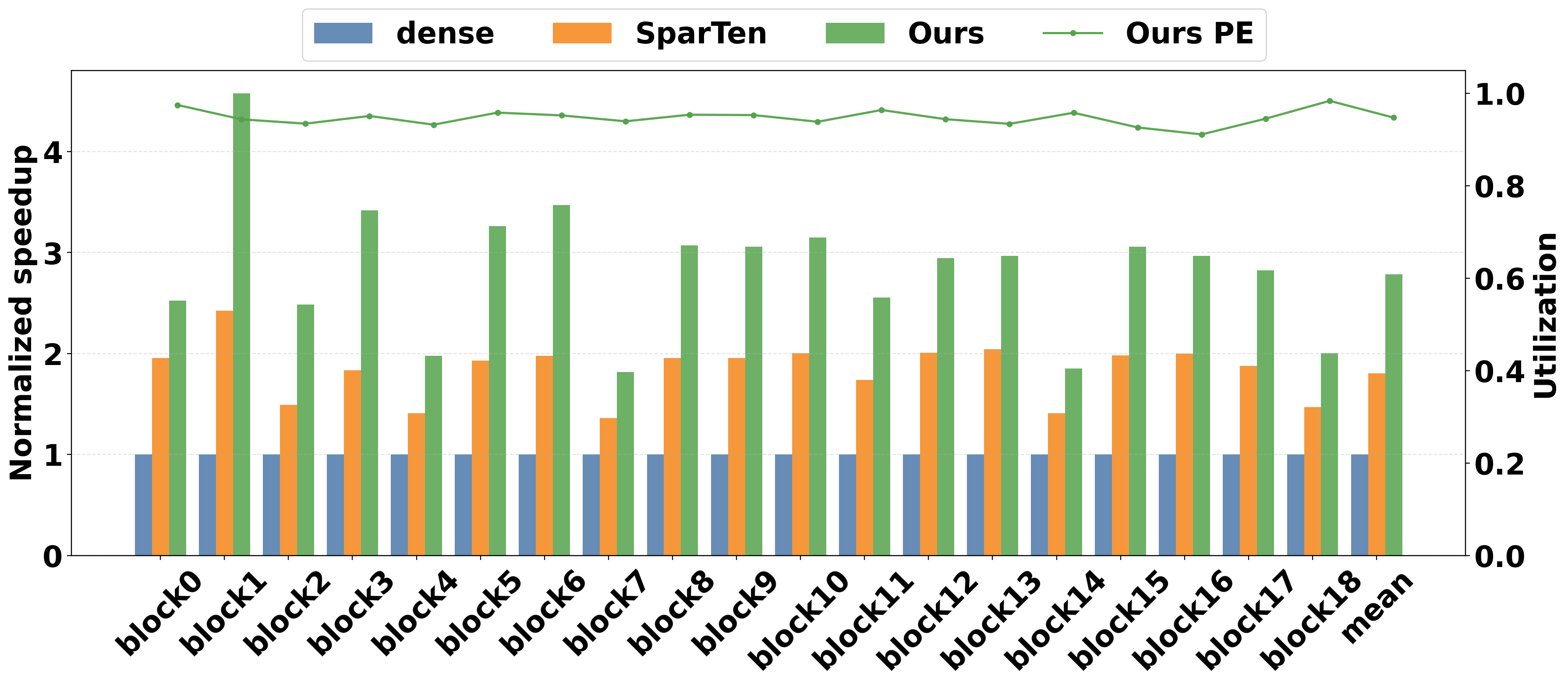}}
    \caption{Performance comparison of our accelerator, SparTen, and a dense accelerator for (a) VGG16 and (b) MobilenetV2.}
    \label{fig:performance}
\end{figure}

Figure \ref{fig:performance} shows the performance comparison of our accelerator, SparTen, and a dense accelerator for VGG16 and MobilenetV2, along with our PE utilization. For VGG16, our accelerator achieves about 6.5$\times$ speedup compared to the dense accelerator and about 1.5$\times$ speedup compared to SparTen. For MobilenetV2, our accelerator achieves about 2.8$\times$ speedup compared to the dense accelerator and about 1.7$\times$ speedup compared to SparTen. Since we designed our hardware configuration to be smaller to meet the needs of edge computing, factors such as SRAM word width, number of banks, number of MAC units, and bitmap size processed by each PE limit the overall speedup. However, we can still maintain very high PE utilization, averaging above 90\%. These results still indicate that our accelerator can more efficiently utilize sparsity than SparTen under limited resources, especially for models with higher sparsity.

Table \ref{tab:comparison} compares our accelerator with other sparse CNN accelerators. While our accelerator has fewer multipliers and smaller core area, it demonstrates excellent energy efficiency, especially on MobilenetV2, achieving 5435-8152 frames/J.

\begin{table}[]
\caption{Comparison with other sparse CNN accelerators.}
\label{tab:comparison}
\begin{center}
\begin{tabular}{|ll|c|c|c|}
\hline
\multicolumn{2}{|c|}{}                                                       & Eyeriss v2~\cite{chen2019eyeriss} & GoSPA~\cite{deng2021gospa}  & Ours          \\ \hline
\multicolumn{2}{|l|}{Technology}                                             & 28nm       & 28nm   & 16nm          \\ \hline
\multicolumn{2}{|l|}{Bit Width}                                              & 8b         & 8b     & 8b            \\ \hline
\multicolumn{2}{|l|}{Number of   Multipliers}                                & 384        & 128    & 64            \\ \hline
\multicolumn{2}{|l|}{Clock Frequency   (MHz)}                                & 464        & 500    & 200-400       \\ \hline
\multicolumn{2}{|l|}{Core Area (mm$^2$)}                                        & 4.08       & 1.76   & 0.5+1.7(SRAM) \\ \hline
\multicolumn{1}{|l|}{\multirow{2}{*}{\begin{tabular}[c]{@{}l@{}}Energy   Efficiency \\ (frames/J)\end{tabular}}} & VGG  & N/A        & 312    & 388-583       \\ \cline{2-5} 
\multicolumn{1}{|l|}{}                                                & MBV2 & N/A        & N/A    & 5435-8152     \\ \hline
\multicolumn{2}{|l|}{Power (mW)}                                             & 461-572    & 89-100 & 12-16         \\ \hline
\end{tabular}
\end{center}
\end{table}

\section{Conclusion}
\label{sec: conclusion}
In this paper, we design and implement an efficient sparse convolution accelerator targeting low-power, low-bandwidth edge IoT vision scenarios. To maximize the benefits of sparse computation under limited hardware resources, we use Bitmap compression format to reduce the volume of data transmission, and propose a multi-dimensional "Idle-First-Task-Assignment" (IFTA) strategy, which effectively resolves the Processing Element (PE) load imbalance issue caused by data sparsity. Meanwhile, considering the characteristics of mobile vision models, we optimize the Depthwise Separable Convolution (DWConv) at both the hardware and dataflow levels. Owing to these architectural optimizations, we achieve $6.5\times$ and $2.8\times$ inference speedups on VGG16 and MobileNetV2 networks under limited resources, respectively, demonstrating higher sparsity utilization efficiency compared to the state-of-the-art SparTen architecture. Based on a complete RTL implementation and 16nm process tape-out, the accelerator chip area is only 0.5 mm$^2$, and the power consumption is controlled within the 12-16 mW range. This demonstrates that our design can deliver excellent performance and energy efficiency in highly constrained edge environments.

\section*{Acknowledgment}

Jingyue Zhuge is funded by Center for Scalable Data Analytics and Artificial Intelligence(ScaDS.AI) and the German Federal Ministry of Education and Research (BMBF) project EVENTS (16ME0733). The authors gratefully acknowledge the computing time made available to them on the high-performance computer at the NHR Center of TU Dresden. 
This center is jointly supported by the Federal Ministry of Education and Research and the state governments participating in the NHR (www.nhr-verein.de/unsere-partner). 

Additional funding by the German Research Foundation (DFG, Deutsche Forschungsgemeinschaft) as part of Germany’s Excellence Strategy – EXC 2050/2 – Project ID 390696704 – Cluster of Excellence “Centre for Tactile Internet with Human-in-the-Loop” (CeTI) of TUD Dresden University of Technology.

\bibliographystyle{IEEEtran}

\bibliography{IEEEtran}

\end{document}